\documentclass[twocolumn,aps,prc,superscriptaddress,showpacs,floatfix,nofootinbib,longbibliography]{revtex4-1}
\usepackage{url}
\usepackage{cancel}
\usepackage[colorlinks,linkcolor=blue,citecolor=blue,filecolor=black,urlcolor=blue]{hyperref}
\usepackage{epsfig,graphics}
\usepackage{graphicx}
\usepackage{dcolumn}
\usepackage{bm}
\usepackage[usenames]{color}
\usepackage{amssymb}
\usepackage{amsmath}
\usepackage{multirow}
\usepackage{float}
\usepackage{harpoon}
\usepackage{MnSymbol}
\usepackage{appendix}
\usepackage{color}
\usepackage{hyperref}
\usepackage{cleveref}

\begin{document}
\title{A comparison study of collisions at relativistic energies involving light nuclei}
\author{Hai-Cheng Wang}
\affiliation{School of Physics Science and Engineering, Tongji University, Shanghai 200092, China}
\author{Song-Jie Li}
\affiliation{School of Physics Science and Engineering, Tongji University, Shanghai 200092, China}
\author{Jun Xu}\email[Correspond to\ ]{junxu@tongji.edu.cn}
\affiliation{School of Physics Science and Engineering, Tongji University, Shanghai 200092, China}
\author{Zhong-Zhou Ren}
\affiliation{School of Physics Science and Engineering, Tongji University, Shanghai 200092, China}
\date{\today}
\begin{abstract}
We present extensive comparisons of $^{16}$O+$^{16}$O collisions at the center-of-mass energy per nucleon pair $\sqrt{s_{NN}}=200$ GeV and $^{208}$Pb+$^{16}$O collisions at $\sqrt{s_{NN}}=68.5$ GeV as well as $^{20}$Ne+$^{20}$Ne collisions at $\sqrt{s_{NN}}=200$ GeV and $^{208}$Pb+$^{20}$Ne collisions at $\sqrt{s_{NN}}=68.5$ GeV based on a multiphase transport (AMPT) model. We recommend measuring the ratio of the elliptic flow to the triangular flow, which shows appreciable sensitivity to the structure of light nuclei as also found in other studies. This is especially so if the observable is measured near the target rapidity in $^{208}$Pb+$^{16}$O or $^{208}$Pb+$^{20}$Ne collisions, as originally found in the present study. Our study serves as a useful reference for understanding the structure effect on observables in collisions involving light nuclei under analysis or on the schedule.
\end{abstract}
\maketitle

\section{Introduction}

Understanding the nucleus structure is a fundamental goal of nuclear physics. Besides traditional methods, relativistic heavy-ion collisions provide an alternative way of extracting the density distribution of colliding nuclei~\cite{Jia:2022ozr}. In the past few years, various probes for nucleus deformation have been proposed and applied in experimental analysis. The basic idea is that the deformation of colliding nuclei may enlarge the anisotropy of the overlap region and enhance the fluctuation of the overlap area in the initial stage of the collision, and this will lead to larger anisotropic flows and stronger transverse momentum fluctuations in the final stage of the collision~\cite{Jia:2021tzt,Jia:2021qyu}. Using this principle, people have successfully extracted the deformation parameters of heavy nuclei such as $^{96}$Ru~\cite{Zhang:2021kxj}, $^{96}$Zr~\cite{Zhang:2021kxj}, $^{197}$Au~\cite{Giacalone:2021udy}, and $^{238}$U~\cite{STAR:2024wgy} as well as the shape of $^{129}$Xe~\cite{Bally:2021qys,ATLAS:2022dov,ALICE:2024nqd}. The focus of the community has now turned to collisions involving light nuclei~\cite{YuanyuanWang:2024sgp,Giacalone:2024luz,Giacalone:2024ixe,Prasad:2024ahm,Lu:2025cni,Zhao:2024feh,Zhang:2024vkh}, which have not only large deformation but possibly different $\alpha$-cluster configurations. For example, $^{16}$O could be of a tetrahedron structure formed by four $\alpha$ clusters~\cite{Freer:2017gip,BIJKER2020103735,Tohsaki:2017hen} with considerable octupole deformation, while $^{20}$Ne could be of a bowling-pin structure consisting of five $\alpha$ clusters~\cite{Zhou:2023vgv,Ropke:2024rlr} with both considerable quadrupole and octupole deformation. The existence of $\alpha$-cluster structure may break the scaling relation between the deformation parameter and specific observables in relativistic heavy-ion collisions such as the anisotropic flows and the transverse momentum fluctuation~\cite{Wang:2024ulq}. While these observables are sensitive to the deformation of colliding nuclei, it is of interest to directly probe the existence of $\alpha$-cluster structure in light nuclei through their collisions at relativistic energies~\cite{Liu:2025zsi}. On the experimental side, besides the $^{16}$O+$^{16}$O (light-light) collisions at top Relativistic Heavy-Ion Collider energy, which is currently under analysis~\cite{Huang:2023viw}, the System for Measuring Overlap with Gas at the Large Hadron Collider beauty experiments enables the study of fixed-target collisions at relativistic energies, and collisions involving light nuclei such as $^{208}$Pb+$^{16}$O and $^{208}$Pb+$^{20}$Ne (heavy-light) collisions are on the schedule (see, e.g., Refs.~\cite{Mariani:2021nwc,LHCb:2022qvj,Aaij_2022,Giacalone:2024ixe}).

It is of interest to compare and understand the different effects of deformation and clustering in light-light collisions and heavy-light collisions on the collision dynamics and final observables. Heavy-light collisions are carried out at a lower center-of-mass collision energy but have a larger size of the initial overlap area, compared to light-light collisions. The event-by-event fluctuation is expected to be stronger in light-light collisions than in heavy-light collisions. Therefore, different correspondence between initial anisotropies and final flows as well as that between the initial overlap size and the final transverse momentum distribution may exist in light-light and heavy-light collisions. While it is difficult to probe directly the existence of $\alpha$-cluster structure in light-light collisions through mid-rapidity observables~\cite{Liu:2025zsi}, it is of interest to investigate whether this is the case in heavy-light collisions. Also, in asymmetric systems such as heavy-light collisions, observables at which rapidity are more sensitive to the structure of light nuclei is a question to be answered. The above will be studied in the present paper based on the AMPT model. We will compare observables in light-light and heavy-light collisions by assuming light nuclei are of a spherical shape, a deformed Woods-Saxon (WS) shape, or with $\alpha$-cluster structure, in order to investigate effects of deformation and $\alpha$-cluster structure in different collision systems.

The rest of the paper is organized as follows. Section~\ref{theo} gives the density distributions of $^{16}$O and $^{20}$Ne and describes briefly the framework of the AMPT model. Section~\ref{results} presents extensive comparisons on the initial anisotropies as well as final anisotropic flows and transverse momentum fluctuations, for different collision systems and with different initial configurations. We give a brief conclusion in Sec.~\ref{summary}.

\section{Theoretical framework}
\label{theo}

The density distributions of $^{16}$O and $^{20}$Ne are obtained based on a Bloch-Brink wave function approach, where empirical nucleon-nucleon interactions are used and the nucleon wave function inside each $\alpha$ cluster is approximated as a Gaussian form. It is assumed that four $\alpha$ clusters form a tetrahedron structure in $^{16}$O and five $\alpha$ clusters form a bowling-pin structure in $^{20}$Ne. The distance parameters in the corresponding $\alpha$-cluster configurations are determined by minimizing the total energy after the angular-momentum projection of the whole wave function. For details of the framework in obtaining the wave functions and the density distributions in $^{16}$O and $^{20}$Ne, we refer the reader to Ref.~\cite{Wang:2024ulq}.

\begin{figure}[ht]
  \centering
  \includegraphics[scale=0.15]{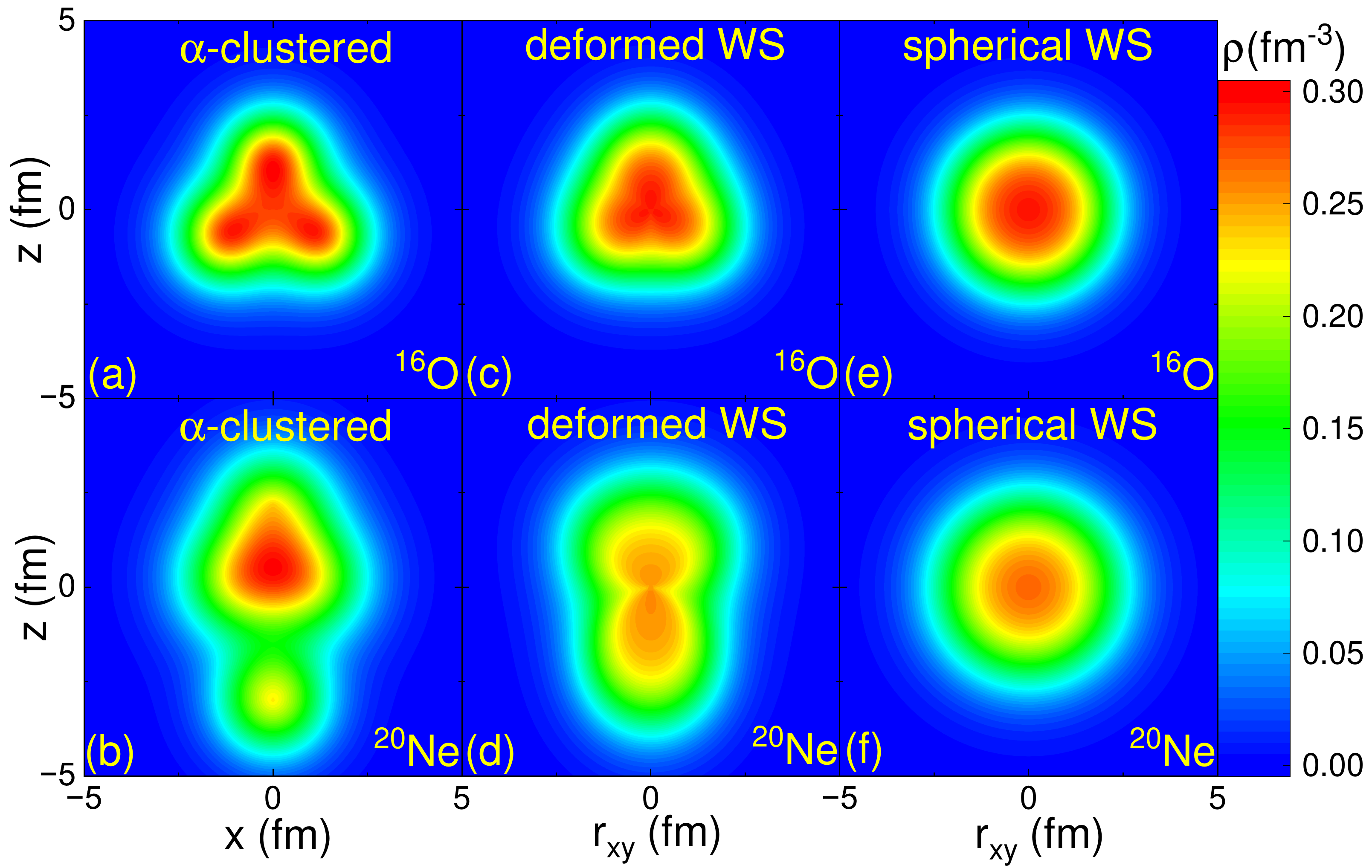}
  \caption{Density contours of $^{16}$O and $^{20}$Ne with $\alpha$-cluster structure (left column), of a deformed WS form (middle column), and of a spherical WS form (right column). $r_{xy}$ represents the radius in x-o-y plane.}
  \label{fig1}
\end{figure}

Figures~\ref{fig1} (a) and (b) display the density contours of $^{16}$O and $^{20}$Ne with $\alpha$-cluster structure obtained from the above framework. In order to distinguish the $\alpha$-cluster effect from the deformation effect on final observables, we have also constructed density distributions of the deformed WS form for both $^{16}$O and $^{20}$Ne as follows
\begin{equation}\label{dfms}
\rho(r,\theta) = \frac{\rho_0}{1+\exp\left\{\frac{r-R_0[1+\beta_2 Y_{2,0}(\theta)+\beta_3 Y_{3,0}(\theta)]}{d}\right\}}.
\end{equation}
In the above, $\rho_0$ is the normalization constant, $R_0$ is the radius parameter, $d$ is the diffuseness parameter, $\beta_{2}$ and $\beta_{3}$ are the deformation parameters, and $Y_{2,0}$ and $Y_{3,0}$ are the spherical harmonics. The deformation parameters $\beta_n$ are determined in such a way that the WS-form density distributions have the same intrinsic multipole moments
\begin{equation}
Q_n = \int \rho(\vec{r}) r^n Y_{n,0}(\theta) d^3 r
\end{equation}
as the density distributions $\rho(\vec{r})$ with $\alpha$-cluster structure. $R_0$ and $d$ are determined in such a way that the deformed WS distributions have the same RMS radii characterized by $\langle r^2 \rangle$ and the same surface diffuseness characterized by $\langle r^4 \rangle$ as those from the density distributions $\rho(\vec{r})$ with $\alpha$-cluster structure, where the $l$th-order moment of $r$ is defined as $\langle r^l \rangle = \int \rho({\vec{r}}) r^l d^3 {r}/\int \rho({\vec{r}}) d^3 {r}$. In this way, the density distributions of $^{16}$O and $^{20}$Ne with the above deformed WS form have the same global shape as the more realistic ones with $\alpha$-cluster structure. The values of $R_0$, $d$, $\beta_2$, and $\beta_3$ in the WS-form density distributions for both $^{16}$O and $^{20}$Ne are listed in Tab.~\ref{T1}. It is seen that the density distribution of $^{16}$O has a considerable $\beta_3$, and that of $^{20}$Ne has both considerable $\beta_2$ and $\beta_3$, as intuitively expected. The resulting density contours for the deformed WS distributions are displayed in Figs.~\ref{fig1} (c) and (d). In order to investigate separately the deformation effect of colliding nuclei on final observables, we have further considered the distributions of the spherical WS form by simply setting $\beta_n=0$ in Eq.~(\ref{dfms}). In this way the nucleus size is the same for the three cases, while the central density in the spherical WS form is higher than that from the spherical Skyrme-Hartree-Fock calculation~\cite{Liu:2023gun}.

\begin{table}[h!]
\centering
\caption{Parameter values for the deformed-WS density distributions of $^{16}$O and $^{20}$Ne.}
\label{T1}
\renewcommand\arraystretch{1.5}
\setlength{\tabcolsep}{3mm}
\begin{tabular}{cccccc}
\hline\hline
 & $R_0$ (fm)& $d$ (fm) & $\beta_2$ & $\beta_3$ \\
\hline
$^{16}$O  & 1.973 & 0.507 & 0 & 0.223 \\
$^{20}$Ne & 2.160 & 0.580 & 0.666 & 0.250 \\
\hline\hline
\end{tabular}
\end{table}

To describe non-equilibrium dynamics in collisions involving light nuclei, transport models are more favored compared to hydrodynamics models. We will compare $^{16}$O+$^{16}$O collisions at $\sqrt{s_{NN}}=200$ GeV under experimental analysis and $^{208}$Pb+$^{16}$O collisions at $\sqrt{s_{NN}}=68.5$ GeV on the schedule, as well as the hypothetical $^{20}$Ne+$^{20}$Ne collisions at $\sqrt{s_{NN}}=200$ GeV and $^{208}$Pb+$^{20}$Ne collisions at $\sqrt{s_{NN}}=68.5$ GeV on the schedule, based on the AMPT model. The density distribution of $^{208}$Pb is set as the empirical spherical WS form in AMPT. The coordinates of participant nucleons in $^{16}$O or $^{20}$Ne are sampled according to the density distributions shown in Fig.~\ref{fig1} with recenterings and random orientations. In the string melting version of the AMPT model used in the present study, hadrons generated by the Heavy-Ion Jet Interacting Generator model~\cite{Wang:1991hta} from collisions of participant nucleons are converted to partons according to the flavor and spin structures of their valence quarks. The momentum spectrum of these initial partons is described by the Lund string fragmentation function
\begin{equation}\label{lund}
f(z) \propto z^{-1} (1-z)^a \exp(-b m_\perp^2/z),
\end{equation}
with $z$ being the light-cone momentum fraction of the produced hadron of transverse mass $m_\perp$ with respect to that of the fragmenting string, and $a$ and $b$ being two parameters. We set $a=0.5$ and $b=0.9$ GeV$^{-2}$ in the present study~\cite{Xu:2011fi,Xu:2011fe}. Sub-nucleon effects~\cite{Mantysaari:2017cni,Welsh:2016siu,Schenke:2014zha,Zheng:2021jrr,Zhao:2021bef,Wang:2025cfu,Wang:2025xzx,Giacalone:2021clp,Alvioli:2013vk,Uzhinsky:2013qga,Alvioli:2017wou}, which may affect the effective area of the nuclear overlap and the initial parton production, are not considered in the present study. The evolution of the partonic phase is described by Zhang's Parton Cascade model~\cite{Zhang:1997ej} including two-body elastic collisions with the differential cross section
\begin{equation}\label{xsection}
\frac{d\sigma}{dt} \approx \frac{9\pi \alpha_s^2}{2(t-\mu^2)^2},
\end{equation}
where $t$ is the standard Mandelstam variable for four-momentum transfer. In the present study, we set the strong coupling constant $\alpha_s$ to be 0.33, and the screening mass $\mu$ to be 3.2 fm$^{-1}$, corresponding to a parton scattering cross section of 1.5 mb~\cite{Xu:2011fi,Xu:2011fe}. After the kinetic freeze-out of these partons, a spatial coalescence model is used to combine quarks or antiquarks into hadrons according to their constituents. A relativistic transport model~\cite{Li:1995pra} including various elastic, inelastic, and decay channels is then used to describe the evolution of the hadronic phase until the kinetic freeze-out of all hadrons. For further details of the AMPT model, we refer the reader to Ref.~\cite{Lin:2004en}. Once the experimental data of, e.g., the charged-particle multiplicity and the anisotropic flows, in the corresponding collision systems are available, we can further calibrate the above parameters used in the AMPT model, which goes beyond the present scope.

\section{Results and discussions}
\label{results}

\begin{figure*}[ht]
  \centering
  \includegraphics[scale=0.15]{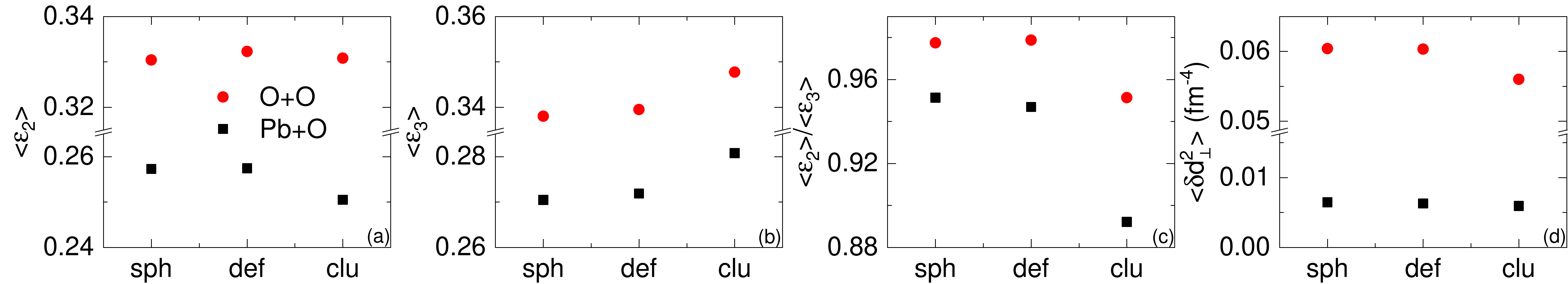}\\
  \includegraphics[scale=0.15]{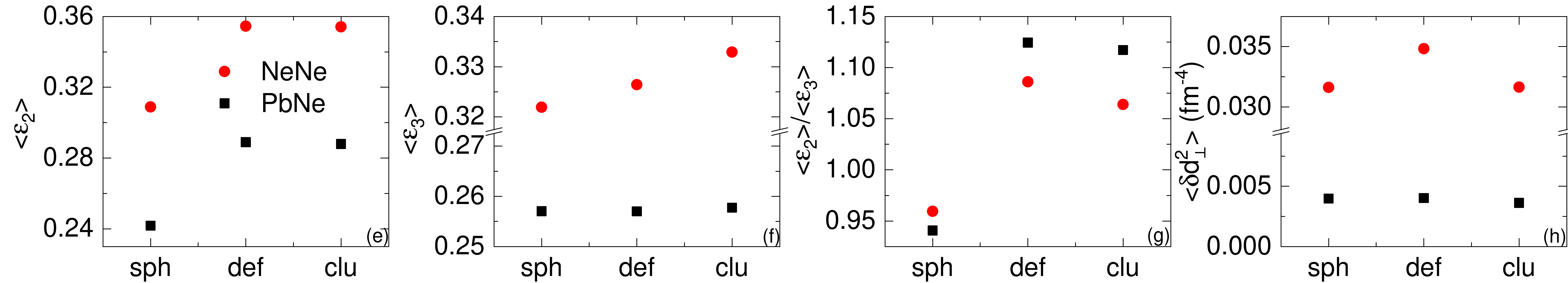}
  \caption{Comparison of the second-order anisotropic coefficient (first column), the third-order anisotropic coefficient (second column), their ratio (third column), and the fluctuation of the overlap's inverse area (fourth column) for initial partons in $^{16}$O+$^{16}$O and $^{208}$Pb+$^{16}$O collisions (upper) as well as in $^{20}$Ne+$^{20}$Ne and $^{208}$Pb+$^{20}$Ne collisions (lower) with the density distributions of $^{16}$O and $^{20}$Ne described by the spherical WS distribution (sph), by the deformed WS distribution (def), and by $\alpha$-cluster structure (clu). }
  \label{fig2}
\end{figure*}

\begin{figure}[ht]
  \centering
  \includegraphics[scale=0.25]{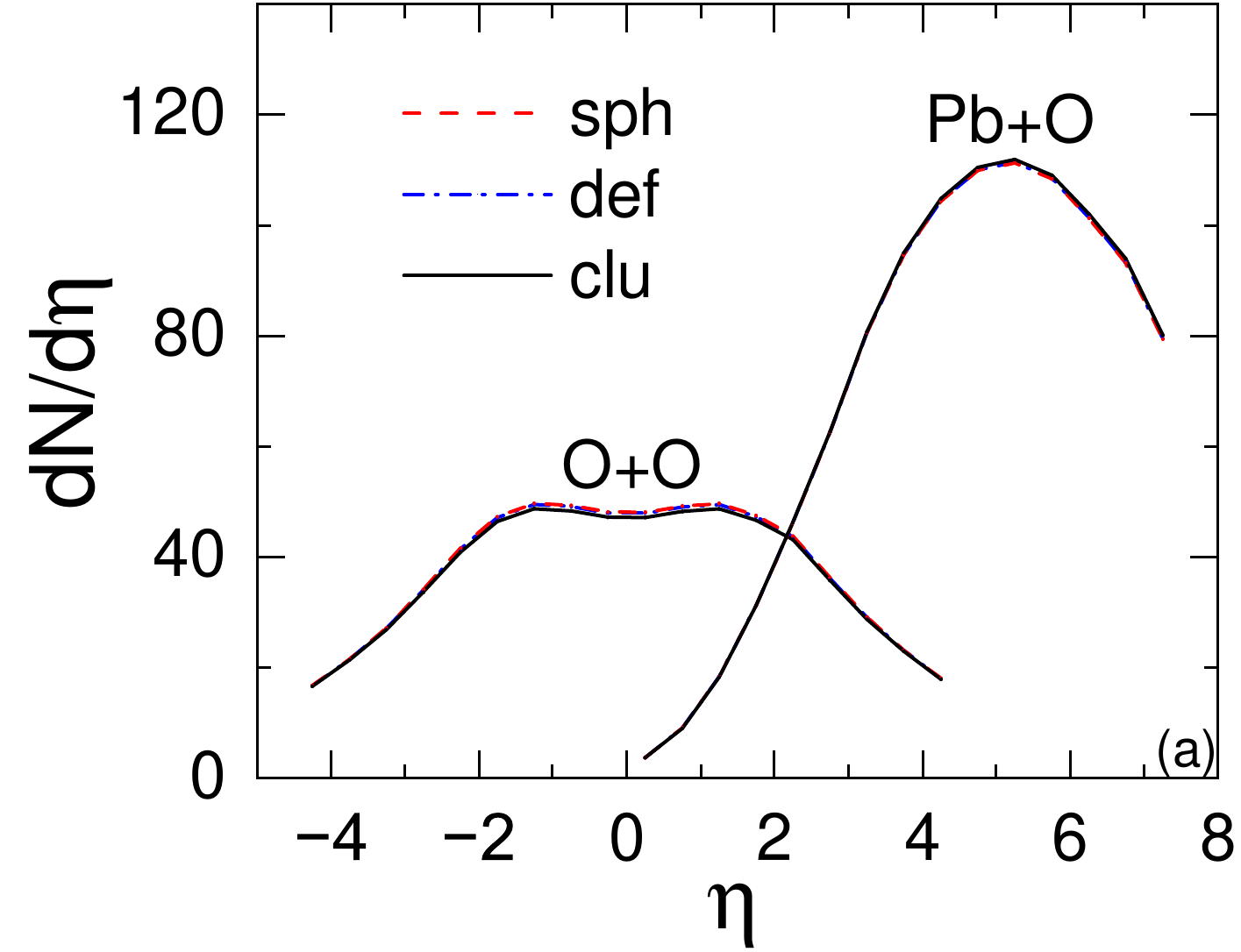}\\
  \includegraphics[scale=0.25]{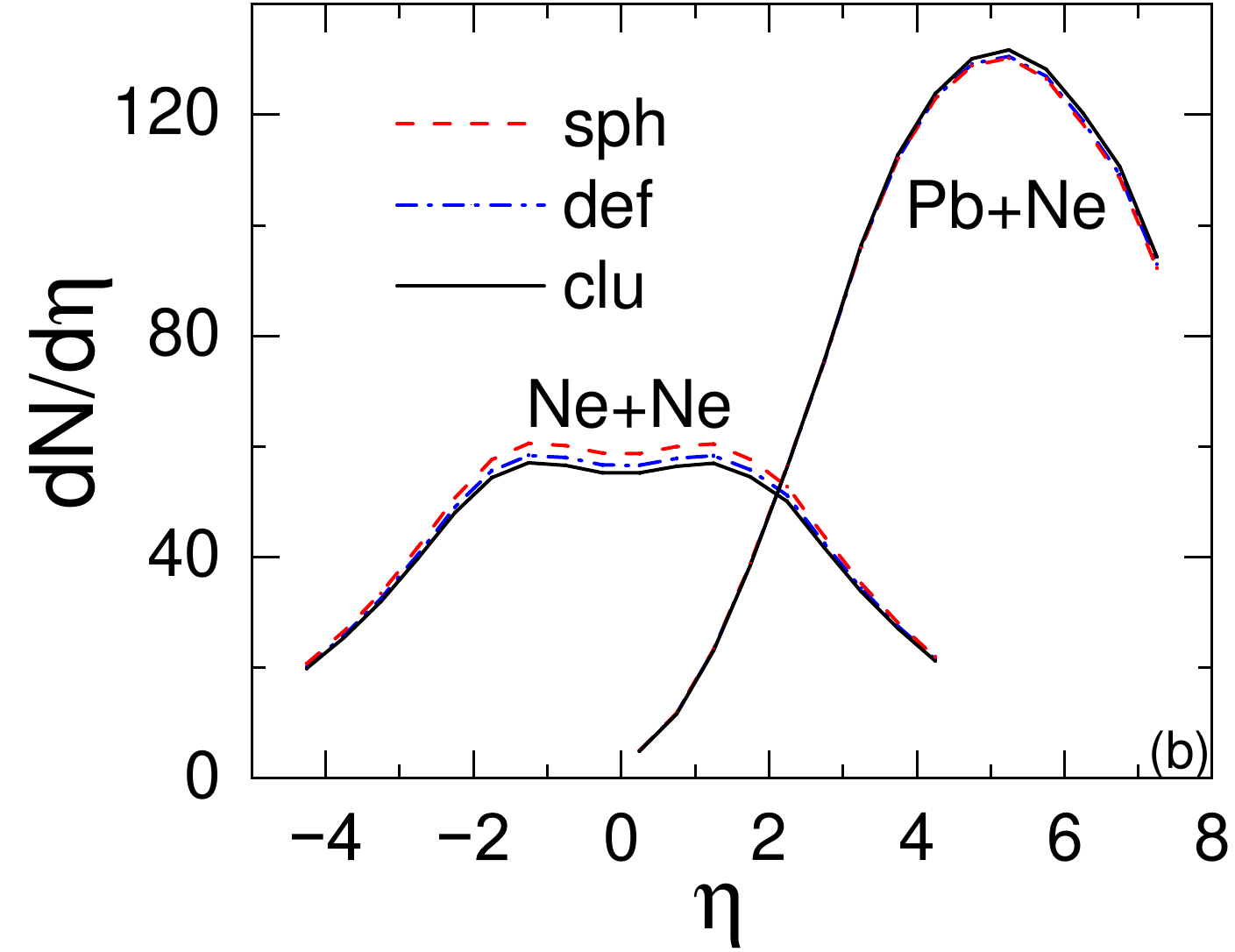}
  \caption{Comparison of final charged-particle pseudorapidity distributions in $^{16}$O+$^{16}$O and $^{208}$Pb+$^{16}$O collisions (upper) as well as in $^{20}$Ne+$^{20}$Ne and $^{208}$Pb+$^{20}$Ne collisions (lower) with the density distributions of $^{16}$O and $^{20}$Ne described by the spherical WS distribution (sph), by the deformed WS distribution (def), and by $\alpha$-cluster structure (clu).}
  \label{fig3}
\end{figure}

\begin{figure*}[ht]
  \centering
  \includegraphics[scale=0.15]{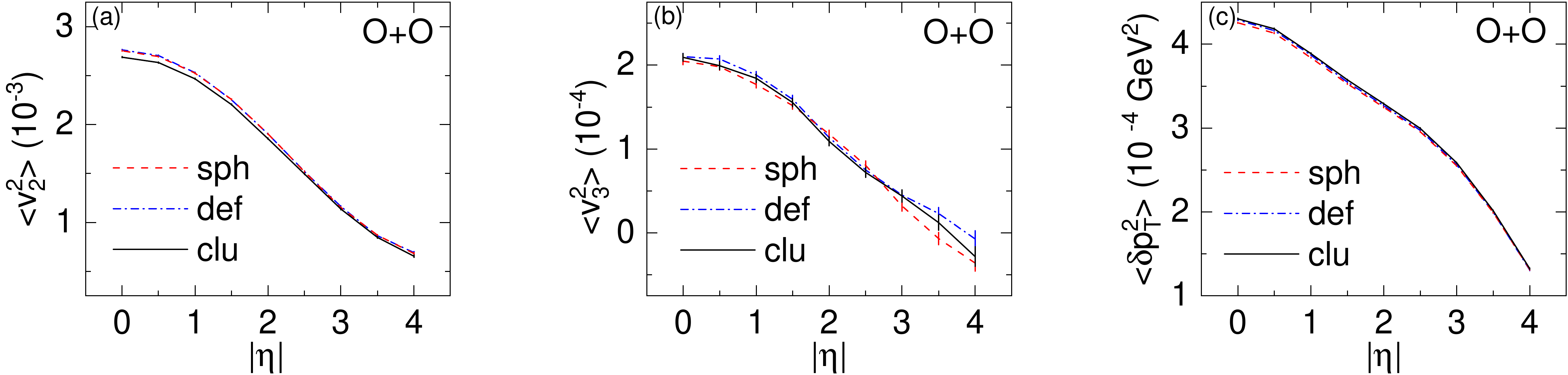}\\
  \includegraphics[scale=0.15]{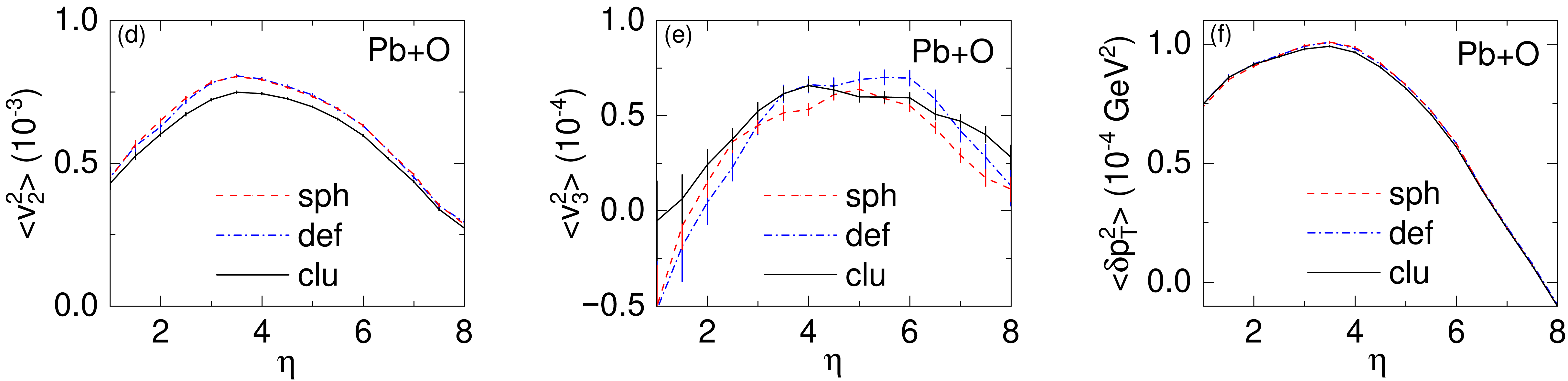}
  \caption{Comparison of pseudorapidity distributions of the elliptic flow (first column), the triangular flow (second column), and the transverse momentum fluctuation (third column) in $^{16}$O+$^{16}$O collisions (upper) and $^{208}$Pb+$^{16}$O collisions (lower) with the density distributions of $^{16}$O described by the spherical WS distribution (sph), by the deformed WS distribution (def), and by $\alpha$-cluster structure (clu).}
  \label{fig4}
\end{figure*}

\begin{figure*}[ht]
  \centering
  \includegraphics[scale=0.15]{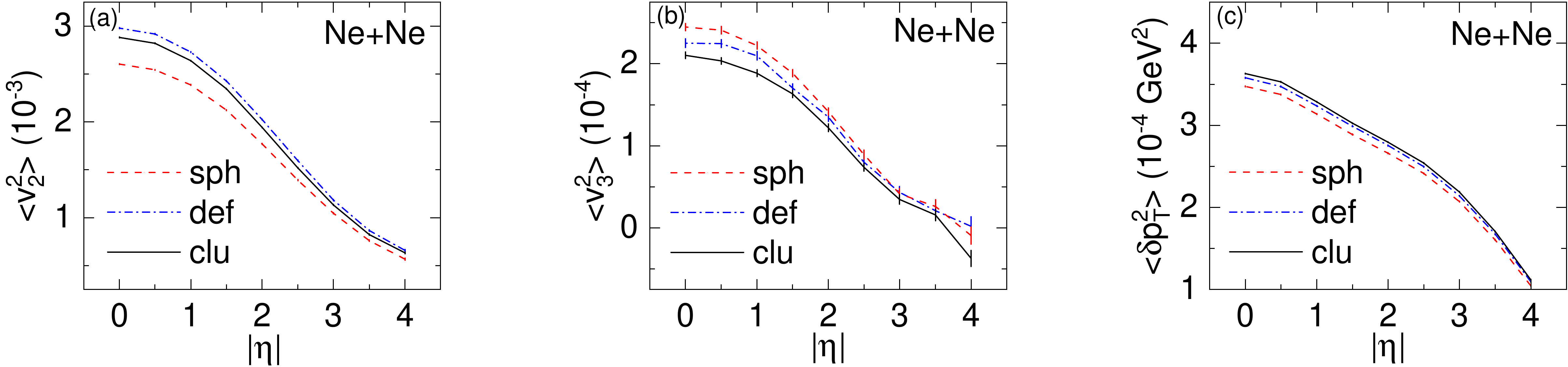}\\
  \includegraphics[scale=0.15]{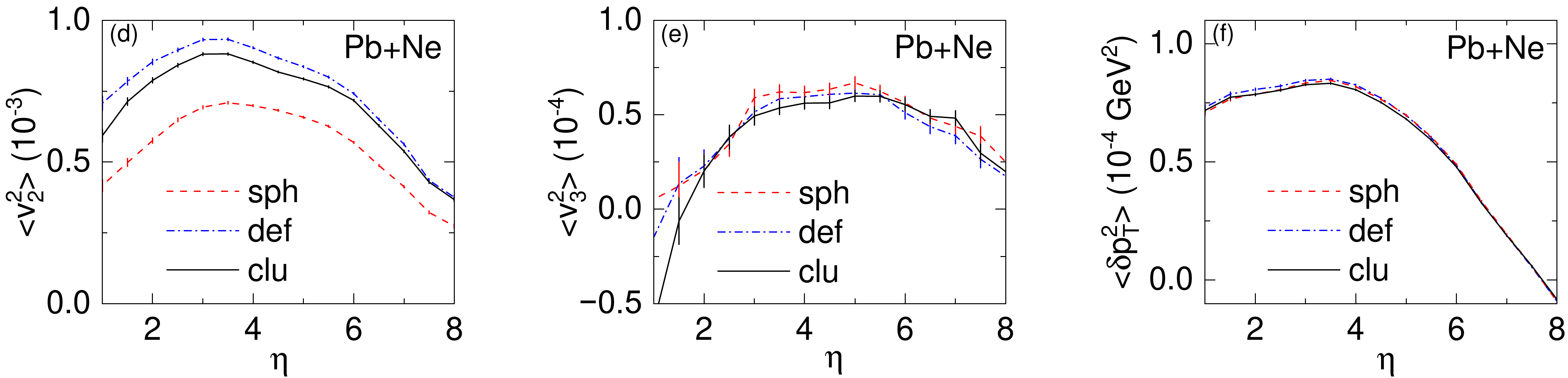}
  \caption{Comparison of pseudorapidity distributions of the elliptic flow (first column), the triangular flow (second column), and the transverse momentum fluctuation (third column) in $^{20}$Ne+$^{20}$Ne collisions (upper) and $^{208}$Pb+$^{20}$Ne collisions (lower) with the density distributions of $^{20}$Ne described by the spherical WS distribution (sph), by the deformed WS distribution (def), and by $\alpha$-cluster structure (clu).}
  \label{fig5}
\end{figure*}

\begin{figure*}[ht]
  \centering
  \includegraphics[scale=0.15]{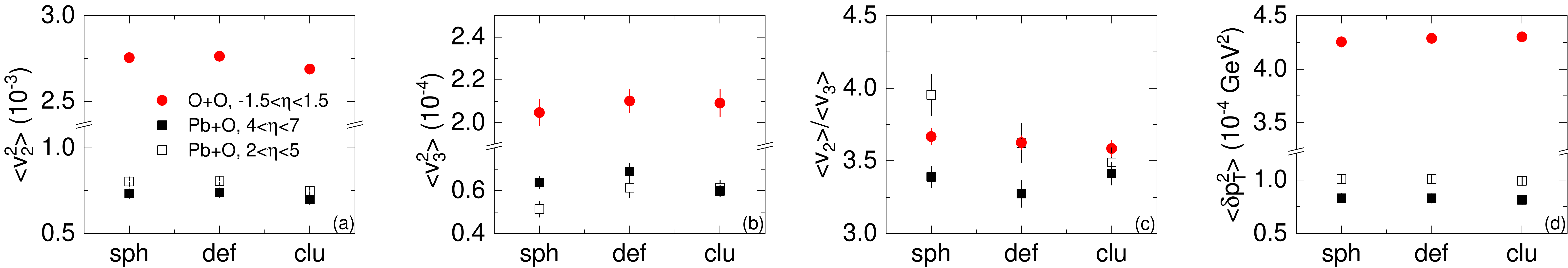}\\
  \includegraphics[scale=0.15]{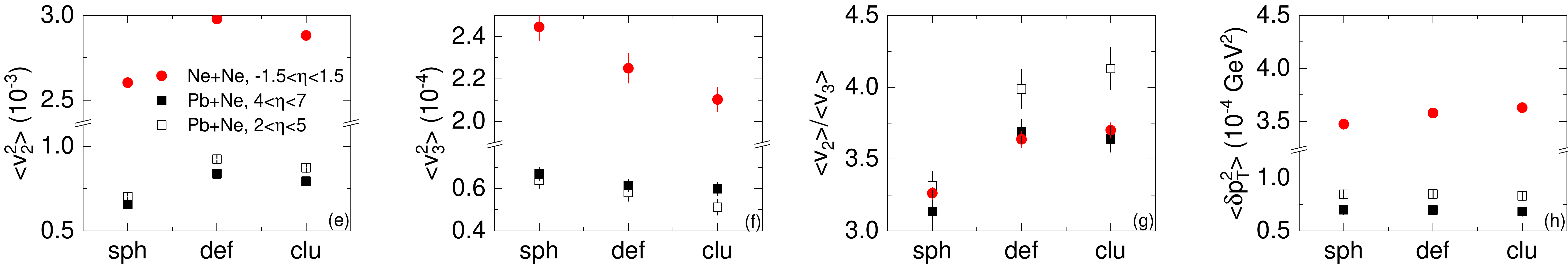}
  \caption{Comparison of the elliptic flow (first column), the triangular flow (second column), their ratio (third column), and the transverse momentum fluctuation (fourth column) in $^{16}$O+$^{16}$O and $^{208}$Pb+$^{16}$O collisions (upper) as well as in $^{20}$Ne+$^{20}$Ne and $^{208}$Pb+$^{20}$Ne collisions (lower) with the density distributions of $^{16}$O and $^{20}$Ne described by the spherical WS distribution (sph), by the deformed WS distribution (def), and by $\alpha$-cluster structure (clu). Results for light-light collisions are calculated at midpseudorapidities ($-1.5<\eta<1.5$), and results for heavy-light collisions are calculated at different forward pseudorapidities ($2<\eta<5$ and $4<\eta<7$).}
  \label{fig6}
\end{figure*}

In this section, we compare extensively observables in $^{16}$O+$^{16}$O collisions at $\sqrt{s_{NN}}=200$ GeV and $^{208}$Pb+$^{16}$O collisions at $\sqrt{s_{NN}}=68.5$ GeV with different initializations of $^{16}$O, and $^{20}$Ne+$^{20}$Ne collisions at $\sqrt{s_{NN}}=200$ GeV and $^{208}$Pb+$^{20}$Ne collisions at $\sqrt{s_{NN}}=68.5$ GeV with different initializations of $^{20}$Ne. We only compare results in central ($0-10\%$) collisions with the maximum multiplicity and the maximum structure effect, with the centrality determined by ordering the charged-particle multiplicities in all events.

We begin by comparing the spatial distributions of partons in the initial stage of different collision systems based on AMPT model calculations in Fig.~\ref{fig2}. Here the $n$th order anisotropic coefficient is calculated from
\begin{equation}
\epsilon_n  =  \frac{ \sqrt{[\sum_i r_{\perp,i}^n \cos(n\phi_i)]^2+[\sum_i r_{\perp,i}^n \sin(n\phi_i)]^2}}{\sum_i r_{\perp,i}^n},
\end{equation}
where $r_{\perp,i}=\sqrt{x_i^2+y_i^2}$ and $\phi_i=\arctan(y_i/x_i)$ are respectively the polar coordinate and the polar angle of the $i$th parton in the transverse plane, and the fluctuation of the overlap's inverse area is defined as
\begin{equation}
\delta d_\perp^2 = (d_\perp - \langle d_\perp \rangle)^2,
\end{equation}
where $d_\perp = 1/\sqrt{\overline{x^2}~\overline{y^2}}$ is the overlap's inverse area~\cite{PhysRevC.102.034905} with $\overline{(...)}$ representing the average over all particles in one event, and $\langle ...\rangle$ represents the average over all events. A smaller number of initial partons and a smaller overlap area are observed in light-light than heavy-light collisions, and this leads to stronger fluctuations and thus systematically larger $\langle \epsilon_n \rangle$ and $\langle \delta d_\perp^2 \rangle$ in light-light collisions. In some cases, the anisotropic coefficients from light-light and heavy-light collision systems show the same sensitivity to the structure of light nuclei. $\langle \epsilon_3 \rangle$ in $^{16}$O+$^{16}$O and $^{208}$Pb+$^{16}$O collisions is rather insensitive to the octupole deformation of $^{16}$O, due to the similar contribution from fluctuations, but is enhanced with $\alpha$-cluster structure in $^{16}$O with a regular tetrahedron configuration. On the other hand, $\langle \epsilon_2 \rangle$ is more sensitive to the quadrupole deformation of colliding nuclei and is enhanced with deformed or $\alpha$-clustered $^{20}$Ne in $^{20}$Ne+$^{20}$Ne and $^{208}$Pb+$^{20}$Ne collisions. However, $\langle \epsilon_n \rangle$ do not necessarily show the same sensitivity to the density distribution of light nuclei in light-light and heavy-light collision systems. It is seen that $\langle \epsilon_2 \rangle$ generated from fluctuations in $^{16}$O+$^{16}$O collisions is insensitive to the density distribution of $^{16}$O, while the $\alpha$-cluster structure in $^{16}$O with a regular tetrahedron configuration reduces slightly fluctuations and thus $\langle \epsilon_2 \rangle$ in $^{208}$Pb+$^{16}$O collisions. $\langle \epsilon_3 \rangle$ is insensitive to the density distribution of $^{20}$Ne in $^{208}$Pb+$^{20}$Ne collisions with more initial partons and larger overlap area, but is enhanced with deformation or $\alpha$-cluster structure in $^{20}$Ne in $^{20}$Ne+$^{20}$Ne collisions due to fewer initial partons and smaller initial overlap area. Consequently, the ratio $\langle \epsilon_2 \rangle/\langle \epsilon_3 \rangle$ is reduced with $\alpha$-cluster structure in $^{16}$O+$^{16}$O and $^{208}$Pb+$^{16}$O collisions, and it is enhanced with deformed or $\alpha$-clustered $^{20}$Ne in $^{20}$Ne+$^{20}$Ne and $^{208}$Pb+$^{20}$Ne collisions. $\langle \delta d_\perp^2 \rangle$ is insensitive to the density distribution of light nuclei in heavy-light collision systems due to the large overlap area, but shows some sensitivity to the density distribution of light nuclei in light-light collision systems.

We now move to the comparison of pseudorapidity distributions of final charged particles in different collision systems shown in Fig.~\ref{fig3}. For the simplicity of the discussion, the pseudorapidity in the following results represents that in the laboratory frame, i.e., $\eta \sim \eta_{lab}$. It is seen that the major multiplicity of charged particles is at the pseudorapidity range of $-1.5<\eta<1.5$ in light-light collisions, while it is at the pseudorapidity range of $4<\eta<7$ in heavy-light collisions, where a significantly higher peak is observed. Although there is no doubt that midrapidity observables are generally used as deformation probes of colliding nuclei in symmetric collision systems, it is tricky to choose the rapidity range for observables in asymmetric collision systems. For heavy-light collision systems, one expects that observables near target rapidities, e.g., $2<\eta<5$, could be more sensitive to the structure of target light nuclei. The multiplicity of charged particles is more sensitive to the density distribution of $^{20}$Ne than to that of $^{16}$O, likely due to the quadrupole deformation of $^{20}$Ne, but this is not adequate to probe the deformation of $^{20}$Ne in experimental analysis.

We then compare extensively pseudorapidity distributions of final-state observables such as the elliptic flow $\langle v_2^2 \rangle$, the triangular flow $\langle v_3^2 \rangle$, and the transverse momentum fluctuation $\langle \delta p_T^2 \rangle$ in light-light and heavy-light collisions for different initial density distributions of light nuclei, and they are calculated according to
\begin{eqnarray}
\langle v_n^2 \rangle &=& \langle \cos [n(\varphi_i-\varphi_j)] \rangle_{i,j}, \label{vn}\\
\langle \delta p_T^2 \rangle &=& \langle (p_{T,i} - \langle \overline{p_T}\rangle) (p_{T,j} - \langle \overline{p_T}\rangle) \rangle_{i,j}. \label{dpt}
\end{eqnarray}
Here $\langle...\rangle_{i,j}$ represents the average over all possible combinations of $i,j$ for all events, and $p_{T,i}=\sqrt{p_{x,i}^2+p_{y,i}^2}$ and $\varphi_i=\arctan(p_{y,i}/p_{x,i})$ are, respectively, the momentum and its polar angle of the $i$th particle in the transverse plane. To calculate $\langle v_n^2 \rangle$ and $\langle \delta p_T^2 \rangle$ at a certain pseudorapidity $\eta$, particles $i$ and $j$ are chosen from the range of $(\eta-1.5, \eta+1.5)$ with a gap of $|\Delta \eta| >1$. Figure~\ref{fig4} compares the results in $^{16}$O+$^{16}$O and $^{208}$Pb+$^{16}$O collisions, and Fig.~\ref{fig5} compares the results in $^{20}$Ne+$^{20}$Ne and $^{208}$Pb+$^{20}$Ne collisions. While in most cases $\langle v_2^2 \rangle$, $\langle v_3^2 \rangle$, and $\langle \delta p_T^2 \rangle$ are strongly correlated with $\langle \epsilon_2 \rangle$, $\langle \epsilon_3 \rangle$, and $\langle \delta d_\perp^2 \rangle$ shown in Fig.~\ref{fig2}, respectively, the detailed behavior depends on the pseudoradipity range and other effects such as particle multiplicities, nonflow, etc. In Fig.~\ref{fig4}, the similar $\langle v_2^2 \rangle$ for different cases in $^{16}$O+$^{16}$O collisions and the smaller $\langle v_2^2 \rangle$ with $\alpha$-cluster structure in $^{208}$Pb+$^{16}$O collisions are consistent with the behaviors of the corresponding $\langle \epsilon_2 \rangle$ shown in Fig.~\ref{fig2} (a). The relative difference in $\langle v_3^2 \rangle$ from different initializations of $^{16}$O depends on the pseudorapidity, and is generally difficult to tell compared with the statistical error. $\langle \delta p_T^2 \rangle$ shows no sensitivity to the initialization of $^{16}$O in both $^{16}$O+$^{16}$O and $^{208}$Pb+$^{16}$O collisions. In Fig.~\ref{fig5}, the larger $\langle v_2^2 \rangle$ from deformed or $\alpha$-clustered $^{20}$Ne in both $^{20}$Ne+$^{20}$Ne and $^{208}$Pb+$^{20}$Ne collisions are consistent with the behaviors of the corresponding $\langle \epsilon_2 \rangle$ shown in Fig.~\ref{fig2} (e). Compared to the spherical case, $\langle v_3^2 \rangle$ slightly decreases after incorporating the deformation or $\alpha$-cluster structure in $^{20}$Ne in both $^{208}$Pb+$^{20}$Ne and $^{20}$Ne+$^{20}$Ne collisions, different from the relative difference in $\langle \epsilon_3 \rangle$ for different scenarios shown in Fig.~\ref{fig2} (f). This is due to the effect of the quadrupole deformation of $^{20}$Ne, which may reduce the triangular flow based on AMPT simulations as shown in Fig. 3 (d) of Ref.~\cite{Wang:2024ulq}. Besides that $\langle \delta p_T^2 \rangle$ is slightly smaller with spherical $^{20}$Ne in $^{20}$Ne+$^{20}$Ne collisions, $\langle \delta p_T^2 \rangle$ shows almost no sensitivity to the initialization of $^{20}$Ne in both $^{20}$Ne+$^{20}$Ne and $^{208}$Pb+$^{20}$Ne collisions. 

Let's then focus on observables at typical pseudorapidity ranges, and the results are shown in Fig.~\ref{fig6}. For light-light collisions, we focus on the pseudorapidity range with the largest multiplicity of final charged particles, i.e., $-1.5<\eta<1.5$. For heavy-light collisions, we focus on the pseudorapidity range with the largest multiplicity of final charged particles, i.e., $4<\eta<7$, and the range near target rapidities, i.e., $2<\eta<5$, at which we expect that the results could be more sensitive to the structure of target light nuclei. Systematically, light-light collisions have stronger anisotropic flows and transverse momentum fluctuations than heavy-light collisions. The behaviors of $\langle v_2^2 \rangle$ are roughly similar to those of $\langle \epsilon_2^2 \rangle$ as shown in Figs.~\ref{fig2} (a) and (e), while the relative $\langle v_3^2 \rangle$ from different initializations of light nuclei are different compared to that in Figs.~\ref{fig2} (b) and (f). This shows that the elliptic flow manifests more clearly the initial geometry while the triangular flow can be affected by other effects in dynamics of collisions involving light nuclei. Taking the ratios of the elliptic flow to the triangular flow in different collision systems shows interesting behaviors. While $\langle v_2 \rangle/\langle v_3 \rangle$ shows almost no sensitivity to the structure of $^{16}$O in the pseudorapidity range with the largest particle multiplicity in both $^{16}$O+$^{16}$O and $^{208}$Pb+$^{16}$O collisions, it shows appreciable sensitivity to the structure of $^{16}$O near target rapidities in $^{208}$Pb+$^{16}$O collisions, with the spherical case giving the largest $\langle v_2 \rangle/\langle v_3 \rangle$ and the deformation or $\alpha$-clustered $^{16}$O leading to smaller $\langle v_2 \rangle/\langle v_3 \rangle$. On the other hand, $\langle v_2 \rangle/\langle v_3 \rangle$ is enhanced by deformation or $\alpha$-cluster structure in $^{20}$Ne in both $^{20}$Ne+$^{20}$Ne and $^{208}$Pb+$^{20}$Ne collisions, and the effect is stronger near target rapidities in $^{208}$Pb+$^{20}$Ne collisions. We note that the centrality dependence of $\langle v_2 \rangle/\langle v_3 \rangle$ has also been shown to be sensitive to the $\alpha$-cluster structure of $^{16}$O in $^{16}$O+$^{16}$O collisions in Refs.~\cite{Prasad:2024ahm,Hu:2025eid}. The transverse momentum fluctuation shows no sensitivity to the initialization of light nuclei. It is seen that these observables are mainly sensitive to the global shape of light nuclei, while it is difficult to distinguish directly the density distribution of light nuclei with and without $\alpha$-cluster structure in both light-light and heavy-light collisions, since the detailed $\alpha$-cluster structure in light nuclei is mostly smeared out by complicated dynamics in the partonic and hadronic phase.

\section{Summary}
\label{summary}

We present extensive comparisons of $^{16}$O+$^{16}$O collisions at $\sqrt{s_{NN}}=200$ GeV and $^{208}$Pb+$^{16}$O collisions at $\sqrt{s_{NN}}=68.5$ GeV as well as $^{20}$Ne+$^{20}$Ne collisions at $\sqrt{s_{NN}}=200$ GeV and $^{208}$Pb+$^{20}$Ne collisions at $\sqrt{s_{NN}}=68.5$ GeV based on the AMPT model, and the main focus is on the sensitivity of observables to the structure of light nuclei. Light-light collisions give systematically stronger anisotropic flows and transverse momentum fluctuations than heavy-light collisions. We found that the ratio of the elliptic flow to triangular flow is a good probe of light nuclei structure as also mentioned in other studies, while the transverse momentum fluctuation shows weak sensitivity. In heavy-light collisions, observables near the target rapidity show stronger sensitivity to the structure of target light nuclei, as found for the first time in the present study. Observables are generally only sensitive to the global shape of colliding nuclei in both light-light and heavy-light collisions, while the detailed $\alpha$-cluster structure in light nuclei can not be distinguished from the constructed deformed Woods-Saxon shape after the complicated dynamics. Our study serves as a useful reference for understanding the structure effect on observables in collisions involving light nuclei under analysis or on the schedule.

\begin{acknowledgments}
This work is supported by the National Key Research and Development Program of China under Grants No. 2023YFA1606701, the National Natural Science Foundation of China under Grant Nos. 12375125, 12035011, and 11975167, and the Fundamental Research Funds for the Central Universities.
\end{acknowledgments}

\bibliography{comparison}
\end{document}